\documentclass{article}
\usepackage{graphicx}
\usepackage[english]{babel}
\usepackage[utf8x]{inputenc}
\usepackage{amsmath}
\usepackage{graphicx}
\usepackage{tikz}  
\usepackage{tikz-3dplot}
\usepackage{amsmath}
\usepackage{hyperref}
\usepackage{pgfplots}
\usepackage{caption}
\usepackage{algorithm2e}
\usepackage{booktabs}
\usepackage{placeins}
\usepackage{pifont}
\usepackage{caption}
\usepackage{booktabs}
\usepackage{makecell}
\usepackage{siunitx}
\usepackage{tcolorbox}
\usepackage{float}
\usepackage{natbib}
\usepackage{changepage}
\usepackage[preprint]{neurips_2020}

\usepackage{hyperref}       % hyperlinks
\usepackage{url}            % simple URL typesetting
\usepackage{booktabs}       % professional-quality tables
\usepackage{amsfonts}       % blackboard math symbols
\usepackage{nicefrac}       % compact symbols for 1/2, etc.
\usepackage{microtype}      % microtypography

\title{Help the Blind See: Assistance for the Visually Impaired through Augmented Acoustic Simulation}

\author{%
  Alexander Mehta\\
  Theia Project\\
  Sunnyvale, CA \\
  \texttt{amehta633@student.fuhsd.org} \\
   \AND
      Ritik Jalisatgi \\
   Theia Project \\
   Sunnyvale, CA \\
   \texttt{ rjalisatgi943@student.fuhsd.org} \\
    \\
}

\begin{document}

\maketitle

\begin{abstract}
 An estimated 253 million people have visual impairments. These visual impairments affect everyday lives, and limit their understanding of the outside world. This can pose a risk to health from falling or collisions. We propose a solution to this through quick and detailed communication of environmental spatial geometry through sound, providing the blind and visually impaired the ability to understand their spatial environment through sound technology. The model consists of fast object detection and 3D environmental mapping, which is communicated through a series of quick sound notes. These sound notes are at different frequencies, pitches, and arrangements in order to precisely communicate the depth and location of points within the environment. Sounds are communicated in the form of musical notes in order to be easily recognizable and distinguishable. A unique algorithm is used to segment objects, providing minimal accuracy loss and improvement from the normal $O(n^2)$ to $O(n)$ (which is significant, as N in point clouds can often be in the range of $10^5$). In testing, we achieved an R-value of 0.866 on detailed objects and an accuracy of 87.5\% on an outdoor scene at night with large amounts of noise. We also provide a supplementary \href{https://www.youtube.com/watch?v=YrfPQbwcvGg
 }{video demo} of our system.
\end{abstract}

\section{Introduction}

An estimated 253 million have some visual impairment, with 36 million completely blind people (\cite{Ackland2017-uo}). Large risks are posed to those with visual impairment -- most notability falling and inability to avoid obstacles due to little spatial understanding (\cite{Steinman2011-kw,de2004different, gillespie2003interventions}). According to \cite{Crews2016}, 46.7\% of blind persons over the age of 65 reported a fall, which can result in death or severe injury. Solutions to these problems have taken shape in various forms (reported on in Prior Work section). Our solution involves using the visually impaired's audio understanding and Pavlovian Conditioning to address this crisis.

\section{Background}

\subsubsection{Motivation}

The idea of echolocation has been commonly known as an attribute of animals such as bats, but the ability to echolocate extends to humans. Particularly, blind or visually impaired humans can possess traits of echolocation according to a small (N=37) preliminary study by \cite{10.3389/fphys.2013.00098}. This is confirmed by \cite{thaler2016echolocation} which finds that echolocation can be a strong alternative to sight for the blind. This is backed up by \cite{10.1371/journal.pone.0020162} who measured brain data to confirm that blind individuals that echolocate effectively often use the same places of their brain stimulated by visual activity. The paper also confirms an important assumption -- echolocation from both external (such as environmental sounds) and internal assistance (such as headphones) has similar accuracy for blind users. Due to this, many devices have been made to assist with echolocation, but none precisely communicate spatial information such as the size of objects near the user.

\subsubsection{Prior Work}

Several systems have been created in order to address blindness through audio techniques.

\textbf{Object Detection and Segmentation} based assistance by \cite{blindsee} addresses this through a real time system to provide users with descriptions of objects that are played at where the objects are located. This is similar to the technique presented in this paper, except our technique relies on properties of sound instead of text to audio descriptions and gives much more detailed information about the user's environmental spatial geometry. Additionally, \cite{blindsee} connects to a large server for computation, while the new system runs offline on a Raspberry Pi 4 and an Intel RealSense Camera, allowing for less constraints on operation in-the-wild.

\textbf{Text-based} approaches have been introduced by \cite{alternative} by using OCR technology in order to describe signs to the blind. This technology can run on a raspberry pi, but fails to account for the overall scene, and non text markers.

This paper presents an innovative approach to enhance spatial communication through sound by utilizing a two-step strategy. The first is encoding spatial geometry data into a series of sounds (Pitch Based Depth Perception), and the second is performing object-segmentation on the spatial geometry data through floodfill to track potentially important areas. It combines the efficiency \cite{alternative} while describing large amounts of information similar to \cite{blindsee}

\section{Materials and Methods}

The method proposed in this paper addresses the challenge of spatial communication through sound by utilizing a two-part approach. The first is mapping spatial geometry data into a series of sounds (Pitch Based Depth Perception), and the second is performing object-segmentation on the spatial geometry data through floodfill to track potentially important areas.

\subsection{Pitch Based Depth Perception}

We address communicating the locations of points in the user's environment through multiple properties of sound: pitch, pan, and index. We communicate the dimension of depth through pitch, with lower pitched notes indicating further depths, then we address the horizontal dimension through panning (the distribution of a sound's volume between the left and right headphone), and finally we address the vertical dimension through the index in which a note was communicated. When used all together, the locations of multiple points within the user's environment may be interpreted which allows the user to mentally map out their spatial environment.

\begin{center}

	\includegraphics[scale=0.3]{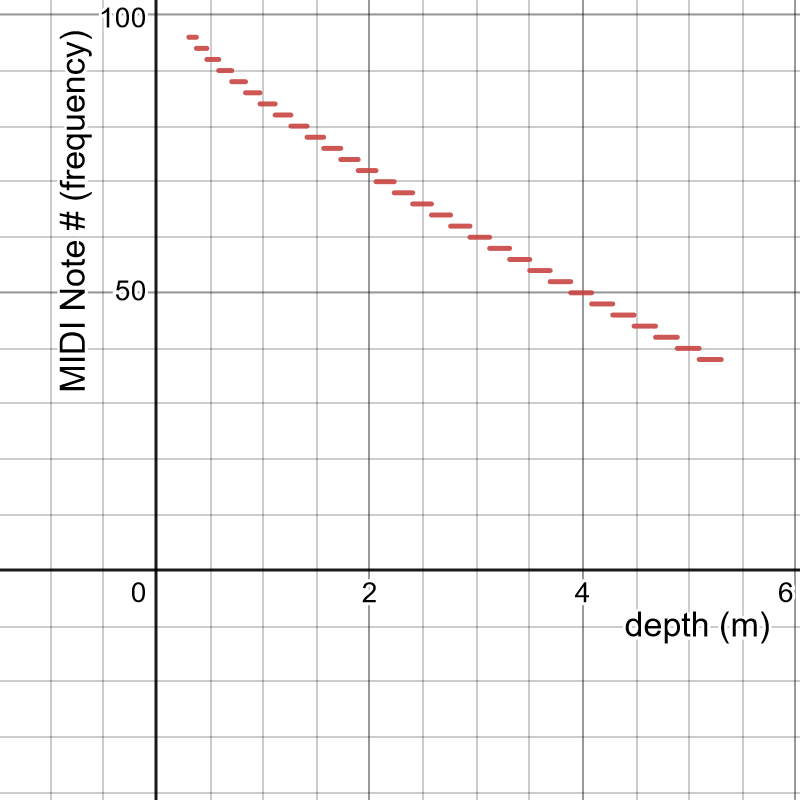}
	
	\href{https://www.desmos.com/calculator/gzovy4wcdu}{Desmos Link: here}
\end{center}
\textbf{Depth to note function:}
\textbf{$96-2\cdot\operatorname{floor}\left(range\cdot\left(\frac{\left(x-start\right)}{\left(start-end\right)}\right)^{0.8}\right)\left\{end\ge x\ge start\right\}$}
We use this function to convert the depth in meters to the pitch of a MIDI note (higher values mean higher pitch).

The function was designed to use the full distinguishable pitch spectrum, and for changes in depth to be more distinguishable at closer values than further values.

	\begin{tikzpicture}[shift={(-3,0)},
		x=3.2cm,y=11cm,
		nodes={text width=3cm},
		image/.style={
			above,
			anchor=south,
			inner xsep=0pt,
		},
		legend/.style={
			below,
			align=left,
			anchor=north,
			inner xsep=0pt,
		},
		]
		
		\draw[black,->,thick,>=latex,line cap=rect]
		(0,0) -- (4.5,0);
		\foreach \Xc in {0,...,4}
		{
			\draw[black,thick]
			(\Xc,0) -- ++(0,5pt);
		}
		\draw (0,0) node[image] {\includegraphics[width=\textwidth]{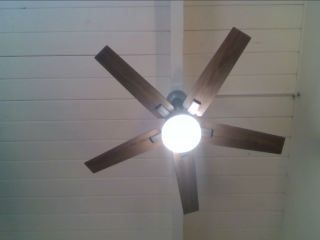}};
		\draw (0,0) node[legend] {Original Image};
		
		\draw (1,0) node[image] {\includegraphics[width=\textwidth]{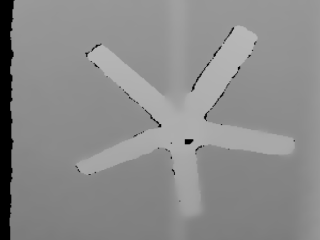}};
		\draw (1,0) node[legend] {Depth Camera Perception};
		
		\draw (2,0) node[image] {\includegraphics[width=\textwidth]{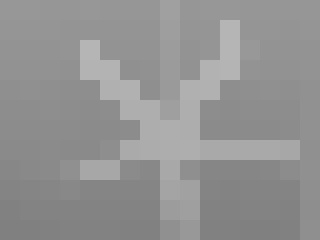}};
		\draw (2,0) node[legend] {Chunking/
			Downsampling};
		
		\draw (3,0) node[image] {\includegraphics[width=\textwidth]{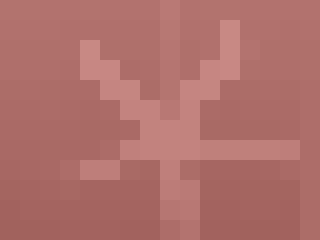}};
		\draw (3,0) node[legend] {Depth to MIDI function applied};

		\draw (4,0) node[image] {\includegraphics[width=\textwidth]{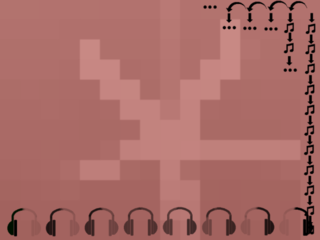}};
		\draw (4,0) node[legend] {Play Sound in specific order};
	\end{tikzpicture}

\hspace{1000pt}
We take the RGBD 2D image array provided by the camera, and then downsample it through nearest neighbor sampling to 16 x 12. Then, using this downsampled RGBD array, we create a MIDI note array by using our depth-note function to convert each depth value in the downsampled array, to a MIDI note value. Each note is also panned between the left and right sides of the user's headphones, with the leftmost value being panned all the way to the left, and the rightmost value all the way to the right. Finally, to communicate this MIDI note array, we play each MIDI note in each column of the array in order from top to bottom, and iterate columns from right to left.

In this study, we employed object detection techniques, as outlined in the subsequent section, to enhance the user's object differentiation ability. The implementation entailed assigning the objects in a scene with unique volume levels. This approach enabled the user to gain a more comprehensive understanding of the environment, through the provision of auditory cues regarding the presence of various objects. It should be noted that this feature was not included in our field demo, as we found that rapid movement could result in disorienting changes due to the volume shifts. These findings will be further explored in the limitations and future work section.

\subsection{Proposed Floodfill Algorithm}
To segment objects within our scene, we use an O(n) floodfill based algorithm that segments the objects based on their relative positions to other points in the pointcloud. We use this floodfill algorithm as a basis to provide the user with better understanding of the objects themselves.

A naive floodfill based segmentation is performed by taking points, calculating distance to other points, and filling the nearest points based on a threshold. Clusters of close points will then constitute an object. Below is a pseudo code version of the naive solution.\\

\begin{algorithm}
	\SetAlgoLined
	\caption{Naive Algorithm}
	
	\KwData{Pointcloud representation of space}
	\KwResult{Segmented Object Pointclouds}
	declare distance threshold\;
	\While{Points in Pointcloud not assigned to object}{
		declare targetPoint\;
		\For{point in Pointcloud}{
			\If{$\sqrt{point^2 + targetPoint^2}<threshold$}{
				place in current object bucket\;
			}
			
		}
		choose new point to perform pointcloud algorithm on\;   	 
		
	}
	
\end{algorithm}

This naive solution performs slowly. Taking $O(n)$ to iterate through N points in a pointcloud. Then for each point, the operation of calculating distances to other points takes $O(n)$, resulting in an $O(n^2)$ algorithm. For RGB-D image point clouds, N can be over $3 * 10^5$ (640 x 480). When N is squared it can result in time complexity of over $9 * 10^{10}$ operations. This is extremely slow, and in echolocation, which is expected to be instant, can be detrimental to the user. Additionally, the operations are slow, as the $\sqrt(x)$ function is slow to compute due to the fact it utilizes Newton's method$^1$ which converges quadratically.

\def\thefootnote{1}\footnotetext{\href{https://encyclopediaofmath.org/index.php?title=Newton_method}{https://encyclopediaofmath.org/index.php?title=Newton\_method}}
In order to solve this, we introduce an algorithm that works in $O(n)$.

We employ a chunking based cache to remove distance calculations. For each point, we round the $x$,$y$, and $z$ to the nearest multiple of $1/n$ for each dimension. These rounded values are considered to be a chunk. This way, a cluster of points in the pointcloud will already be in the same chunk / group. We then create a lookup table where keys are the original coordinates, and the value is the chunk. We can then do the $O(n)$ floodfill that exists when filling a 2D grid, except in 3D. We simply flood the chunks in the same fashion one would flood squares in a grid. This leaves us with groups of chunks, which we then use to determine groups of points by converting the chunks in each group to their stored points.
\begin{algorithm}{
		\SetAlgoLined
		\caption{Optimized Object Detection Algorithm}
		
		\KwData{Pointcloud representation of space}
		\KwResult{Segmented Object Pointclouds}
		declare chunk cache;\
		\For{point in Pointcloud}{
			assign point to specific chunk\;
		}
		declare set of unflooded cache buckets;
		\While{set contains points}{
			declare select point and assign to the first item in set\;
			\If{surrounding cache buckets contain point}{mark those points as part of the current object bucket\; remove from unflooded set\;}

		}
	}
	
\end{algorithm}

This algorithm is able to perform floodfill without calculating relative distances and reducing the complexity of the algorithm to $O(n)$. For the average pointcloud size of $3 * 10^5$. Our algorithm runs in $3*10^5$ operations instead of $9 * 10^{10}$. Additionally, each operation is significantly faster with the removal of the $sqrt(x)$ function. Below is a visual representation.

\begin{tikzpicture}[
	x=3.5cm,
	nodes={text width=3cm},
	image/.style={
		above,
		anchor=south,
		inner xsep=0pt,
	},
	legend/.style={
		below,
		align=left,
		anchor=north,
		inner xsep=0pt,
	},
	]
	\draw[black,->,thick,>=latex,line cap=rect]
	(0,0) -- (3.5,0);
	\foreach \Xc in {0,...,3}
	{
		\draw[black,thick]
		(\Xc,0) -- ++(0,5pt);
	}
	\draw (0,0) node[image] {\includegraphics[width=\textwidth]{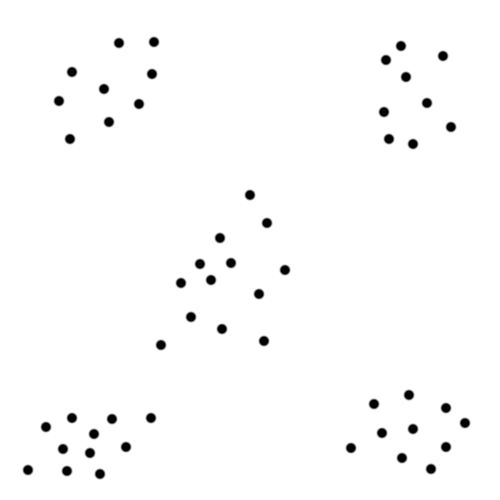}};
	\draw (0,0) node[legend] {Raw Pointcloud Data};
	
	\draw (1,0) node[image] {\includegraphics[width=\textwidth]{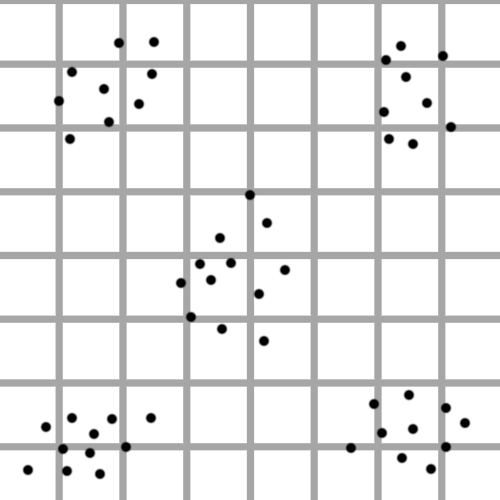}};
	\draw (1,0) node[legend] {3D space is separated into chunks };
	
	\draw (2,0) node[image] {\includegraphics[width=\textwidth]{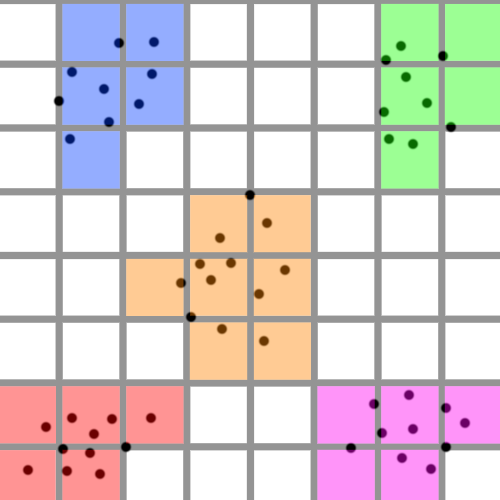}};
	\draw (2,0) node[legend] {Chunks are floodfilled};
	
	\draw (3,0) node[image] {\includegraphics[width=\textwidth]{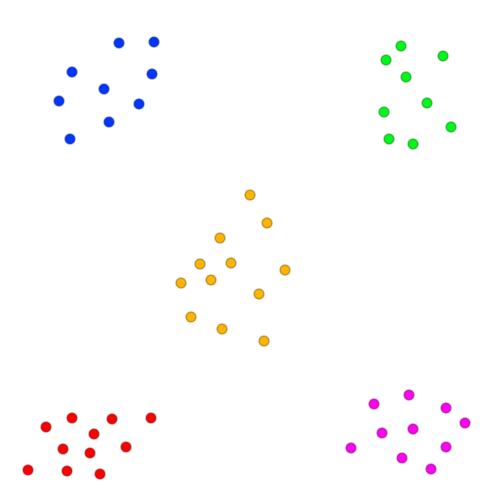}};
	\draw (3,0) node[legend] {Points assigned to object that was assigned to respective chunk};
\end{tikzpicture}

\subsection{Implementation}
We implement our model in both C\# with Unity and Python with PyGame. We use the Unity implementation for standardized tests and the lightweight python version for in the wild inference.

The PyGame version for in-the-wild demos and inference doesn't use object detection, as users in a moving scene may feel disoriented by the fast paced change of objects if too many objects enter and exit a scene (discussed in the limitations section). Additionally, the Unity implementation takes use of Unity's 3D audio protocol in order to effectively convey to the user where a sound comes from. A limitation of this algorithm is headphones often don't provide high quality 3D audio, specifically the ability to distinguish the height of sounds is lacking.

\subsection{Physical System}

To receive spatial input, we use an Intel Realsense D415 depth camera which provides depth information with low compute resources. To run our model portably, we use a Raspberry Pi 4+ ARM based SoC running Ubuntu 22.01. We fit our model onto the user like a headlamp, with the camera being strapped to the users forehead, and computations being ran on a Raspberry Pi 4+ housed in a 3D case which is strapped to the back of the user's head. Power banks of all sizes may be used to power the device, and can be swapped out for a smaller or larger one if necessary.

%insert image of the system on head and on desk
\begin{center}
	\includegraphics[scale=0.7]{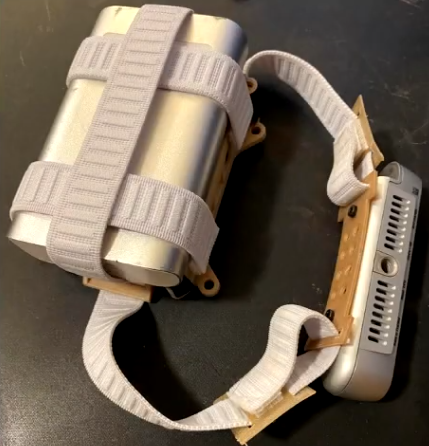}
\end{center}

\subsection{Experiments}

\subsubsection{Object Segmentation Evaluation}

In order to evaluate the object segmentation accuracy of the proposed algorithm in real-world scenarios, we evaluated the performance of segmenting basic elementary shapes from one another. The accuracy of the proposed $O(n)$ algorithm was tested and compared to a $O(n^2)$ baseline algorithm. The running time was measured in milliseconds (ms) and total basic compute operations, and the correct number of segmented objects was evaluated by comparing the difference between the actual and expected objects in the scene using Pearson's Correlation Coefficient, which allows us to measure effectiveness even if the amount of objects detected is not accurate. For example, if the algorithm detected 5 objects instead of 6, the 5 detected would still count towards the score. Additionally, we use a percent accuracy metric to measure the number of times the algorithm ran perfectly. This real world, in-the-wild evaluation method provides a comprehensive assessment of the algorithm's performance in real-world scenarios compared to the base algorithm.

\textbf{Pearson's Correlation Coefficient (referenced to as the R-value)}

\[PCC = \frac{{}\sum_{i=1}^{n} (x_i - \overline{x})(y_i - \overline{y})}
{\sqrt{\sum_{i=1}^{n} (x_i - \overline{x})^2(y_i - \overline{y})^2}} \]

\subsubsection{Pitch Based Depth Perception}

Due to the interpretive nature of the visual to audio based algorithm, we provide in-the-wild demonstrations to provide a comprehensive evaluation of the proposed approach in real-world scenarios (\href{https://www.youtube.com/watch?v=YrfPQbwcvGg}{link here}).

\section{Results}

\subsubsection{Object Detection Results}
We evaluated our model on the Intel RealSense in-the-wild outdoor pointcloud set. We find that the dataset accurately depicted the scene, while having minor noise issues as times when light conditions were unideal.

\begin{center}
	
	\includegraphics[height=3cm]{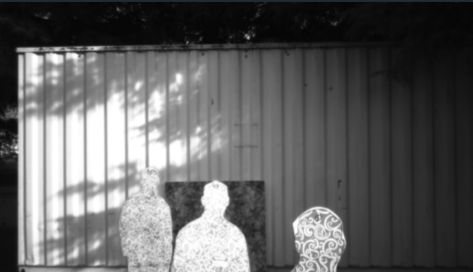}
	\quad
	\includegraphics[height=3cm]{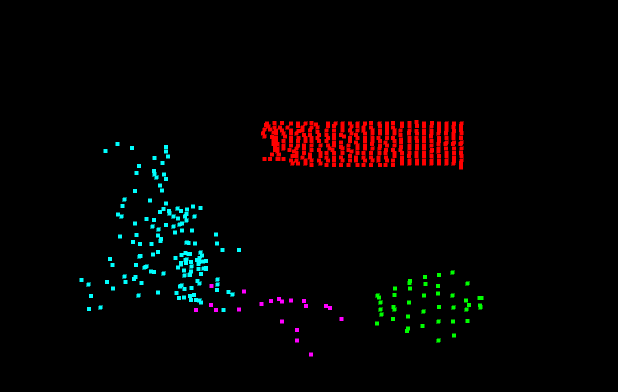}
	
\end{center}

Black and White Image captured by depth camera (left) and our Unity object representation (right). The algorithm detected the correct amount of items in the correct places. Specifically the 3 human cutouts are seen from left to right and the wall behind.

\begin{center}
	
	\includegraphics[height=3cm]{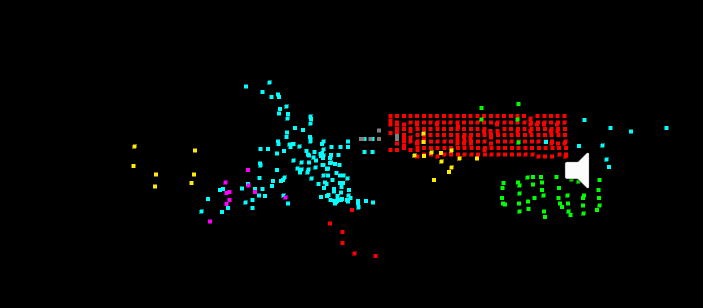}
	
\end{center}

As seen, our algorithm falls short when it comes to small artifacts, believing small artifacts are actually objects.

We evaluated the performance of our algorithm using pointcloud files provided by Intel, containing up to 30 seconds of pointcloud data per file. Additionally, we also tested the algorithm on object files from the dataset presented in the paper by \cite{objectdataset}. Results were collected from three distinct datasets: an outdoor scene captured by a D415 RealSense camera, known for its ability to generate large and accurate pointclouds; an indoor scene captured by a SR300, a lower quality depth camera; and the object dataset by \cite{objectdataset}, which includes high-quality pointclouds of various scenes, not recorded using RealSense technology.
\begin{center}
	\textit{Object Detection Results}
	\centering
	\begin{tabular}{ || c c c c || }
		\hline
		Data Description & Pearsons R  & Accuracy (\%) & Type \\ [0.5ex]
		\hline
		\hline
		Outdoor Scene &  N/a  & 87.5\% & High Quality RealSense with IR \\
		\hline
		Ball & 0.316 & 50\% & Low Quality RealSense \\
		\hline
		
		\cite{objectdataset}  & 0.866 & 12.5\% & High Object Count\\

		\hline
	\end{tabular}
	\textit{Note that Pearsons R is considered N/a if the object count doesn't change}
\end{center}

\subsubsection{Pitch Based Depth Perception}

The present study offers a demonstration of Pitched Based Depth Perception, which can be accessed through the following link: \href{https://www.youtube.com/watch?v=YrfPQbwcvGg}{https://www.youtube.com/watch?v=YrfPQbwcvGg}. It is widely acknowledged that pitch perception is a variable trait among individuals and can be improved through training (\cite{pmid31550277}). As a result, it is not possible to accurately evaluate the effectiveness of this approach to depth perception without conducting a comprehensive evaluation involving a large sample of human participants. This will be tested in future study, as noted in the future work section.

\begin{center}
	
	\includegraphics[height=3cm]{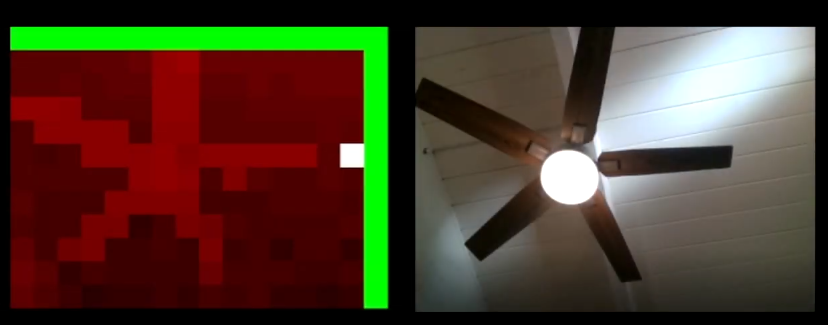}
	
	\textit{    Visualization of pitch based algorithm from \href{https://www.youtube.com/watch?v=YrfPQbwcvGg}{video demo}, where the lighter the color, the greater the pitch.}
	
\end{center}

\begin{center}
	
	\includegraphics[height=3cm]{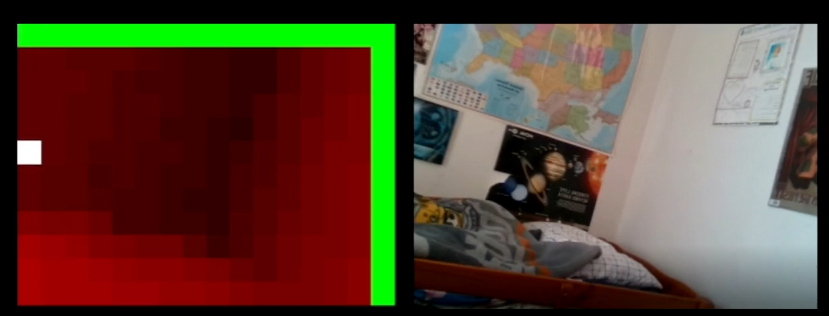}
	
	\textit{More noisy, complex, example}
	
\end{center}

\section{Discussion}

The present study demonstrates that the accuracy of our algorithm in dark outdoor scenes is high and remains robust in the presence of large artifacts, with an accuracy of 87.5\%. This is of particular significance for individuals with visual impairments, as navigating dark environments can be particularly challenging. Furthermore, research has shown that partial blindness is more prevalent during night time hours \cite{discussion_a}. The high level of accuracy achieved in this study may be attributed to the use of high-quality point clouds, suggesting that the algorithm benefits from a greater amount of data. These findings provide motivation for future research to invest in higher quality materials in real-world environments in order to achieve optimal results.

The results of the present study indicate that the accuracy of the algorithm on the ball dataset is relatively low, as evidenced by the R-value of 0.316 and an accuracy rate of 50\%. These findings are consistent with previous claims that the performance of the algorithm is directly correlated with the quality of the data used. Specifically, the use of a lower quality realsense camera appears to have led to a reduction in accuracy. This serves as further evidence of the importance of utilizing high-quality data in order to achieve optimal performance with this algorithm.

The results of the present study indicate that the dataset provided by \cite{objectdataset} yielded high R-value results but relatively low accuracy results. This is a reasonable outcome, given that the dataset used in this study included a larger number of objects in a scene compared to previous datasets, with a size range of $10^1$ to $10^2$. While it may be unrealistic to expect perfect recall in such a scenario, the high R-value values suggest that the algorithm demonstrates a high level of precision, even if the overall accuracy is not perfect.

\section{Conclusion}

In this study, we proposed a novel approach for providing audio-based assistance to individuals with visual impairments. Our approach comprises of two main algorithms, namely a depth perception algorithm and a chunk-based floodfill algorithm. The depth perception algorithm allows the visually impaired to perceive object depth and location in three dimensions through the use of different pitches of sounds played through headphones. The second algorithm is a chunk based floodfilling algorithm. We find that this algorithm improves from the base time complexity of $O(N^2)$ to $O(N)$ through our caching mechanism. We find this algorithm performed well in scenes, but falls short in low quality pointclouds. Overall, we find a large research opportunity in audio based scene understanding to help those who are visually impaired.

\subsubsection{Limits, Implications, and Future Work}

Our findings indicate that the performance of the algorithm is limited by the quality of the depth camera on which it is run. As such, future research should focus on developing the algorithm's robustness to lower quality environments. Additionally, it should be noted that the proposed algorithm is unique in nature and therefore, it is not straightforward to establish a benchmark for comparison with other algorithms in the field. Specifically, the algorithm is exclusively focused on object detection for non-visual, audio-based understanding, making it difficult to determine the required level of accuracy for effective translation from visual to audio-based understanding.

Our evaluation revealed that the depth based perception algorithm functioned as intended, however, it has the major implication of requiring the user to possess a sufficient level of auditory acuity to comprehend the range of pitches employed by the algorithm. In order to fully evaluate the effectiveness of this approach, future research should include human participant testing. Specifically, this testing should focus on assessing the users' ability to comprehend visual scenes through their hearing, in order to determine the minimum level of auditory acuity required for the effective use of this algorithm.

We find that different audio based cue algorithms for depth perception are limited by headphone designs. Due to the lackluster quality of 3D spatial audio headphones, the ability to best convey the scene is hindered by current audio technology. Future improvements to headphone technology will lead to a similar improvement in the abilities of our algorithm to distinguish different directions. Specifically, the ability to differentiate sounds vertically would allow the algorithm to be extended to understanding where sounds were being played from in all 3 dimensions without overwhelming the user with too many sound changes.

In this study, we proposed a solution for providing audio-based assistance to individuals with visual impairments, who constitute an estimated 253 million people worldwide. These visual impairments can significantly affect everyday lives, limiting their understanding of the outside world and posing a risk to their health from falling or collisions. Our solution aims to enhance the mobility and independence of visually impaired individuals by providing quick and detailed communication of environmental spatial geometry through sound. The proposed model consists of fast object detection and 3D environmental mapping, which is communicated through a series of quick sound notes that convey the depth and location of points within the environment. The sounds are communicated in the form of musical notes to make them easily recognizable and distinguishable. A unique algorithm was used to segment objects, resulting in minimal accuracy loss and significant improvement in computational efficiency from the standard $O(n^2)$ to $O(n)$. In testing, we achieved an R-value of 0.866 on detailed objects and an accuracy of 87.5\% in an outdoor scene at night with large amounts of noise. The results of this study demonstrate the potential of audio-based assistance in augmenting the mobility and independence of visually impaired individuals.

\medskip

\small
\bibliography{BibliographyFile}

\end{document}